\documentclass[10pt]{article}

\usepackage{graphicx}
\usepackage{amsmath,amssymb}
\usepackage{bm}
\usepackage{siunitx}
\usepackage{hyperref}
\usepackage[numbers]{natbib}

\title{Modal theory and near-field energy-flow topology in enhanced transmission through subwavelength apertures}

\author{
MA Ortiz-Ferreyro,
J. Sumaya-Martinez,
A. Esquivel-Navarrete\\
\small Department of Physics, Universidad Autonoma del Estado de Mexico, Toluca, Mexico\\
\small \texttt{jsm@uaemex.mx}
}

\date{}

\begin{document}
\maketitle

\begin{abstract}
Enhanced optical transmission through subwavelength apertures is commonly described within a modal framework, where guided modes excited inside the aperture interfere along its length. In this work, we analyze this phenomenon from the perspective of near-field energy transport, using a modal formulation for the two fundamental polarizations, transverse electric (TE) and transverse magnetic (TM). Focusing on subwavelength slits and channels in a perfect conductor, we examine the time-averaged Poynting vector in the vicinity of the aperture as the wavelength is varied across resonance. We show that resonant transmission is accompanied by a pronounced reorganization of the energy-flow field, characterized by vortical circulation, saddle-type stagnation points, localized backflow regions, and efficient energy funneling into the aperture. These features are closely correlated with strong phase gradients and phase singularities of the underlying modal fields. The results can be interpreted within the framework of singular optics applied to the Poynting vector field, providing a physically transparent picture of enhanced transmission as a consequence of modal interference and near-field energy-flow topology. This approach offers a unified description of enhanced transmission in slits and channels that does not rely on plasmonic effects and is applicable beyond specific geometries.
\end{abstract}

\section{Introduction}
Enhanced optical transmission through apertures whose transverse dimensions are smaller than the wavelength has been extensively studied in subwavelength optics, beginning with the extraordinary transmission reported in periodic hole arrays \cite{EbbesenNature1998,MartinMorenoPRL2001}. While early interpretations emphasized surface plasmon polaritons in real metals, it is now well established that large transmission can also occur in perfect conductors and in regimes where plasmonic modes are absent. In such cases, a natural and widely used description is provided by modal theory, in which the aperture behaves as an open waveguide supporting a discrete set of TE and TM modes.

Within this modal framework, resonant transmission is commonly interpreted as a Fabry--P\'erot--like interference of guided modes along the thickness or length of the aperture \cite{Takakura,PortoPRL1999,LalannePRL,GarciaVidalRMP}. This description successfully accounts for resonance conditions, polarization selectivity, and the dependence on geometry for subwavelength slits and channels.

Most modal analyses focus on transmission spectra and far-field observables. Here we complement the modal picture by explicitly analyzing how modal excitation reorganizes electromagnetic energy transport in the near field. We compute the time-averaged Poynting vector and show that the resonant regime is accompanied by a qualitative reconfiguration of the energy-flow field, including stagnation points, saddle points, vortical circulation, and localized backflow. We further connect these flow features with phase structure and phase singularities in the underlying complex fields \cite{NyeBerry,DennisProgOpt,BerryDennis2001}. Here, the term topology refers to the qualitative organization of the energy-flow vector field (streamlines, stagnation points, vortices, and separatrices), rather than to topological invariants in band theory.

\section{Modal formulation for TE and TM polarizations}
We consider a two-dimensional aperture in a perfectly conducting screen, invariant along $y$, with cross-section in the $x$--$z$ plane. The incident field is monochromatic with angular frequency $\omega$ and free-space wavenumber $k_0=\omega/c$.

\subsection{Waveguide modes inside the aperture}
Inside the aperture, fields are expanded in guided modes of the corresponding waveguide \cite{CollinBook}. For each polarization, the field can be written as a superposition of forward- and backward-propagating modes with propagation constants $\beta_m$. For subwavelength widths, the fundamental mode dominates. In particular: (i) for TM polarization (nonzero $H_y$), the fundamental mode has no cutoff and can propagate for arbitrarily long wavelengths; (ii) for TE polarization (nonzero $E_y$), the fundamental mode exhibits a cutoff wavelength and becomes evanescent below cutoff. These modal properties explain the strong polarization selectivity observed in the deep-subwavelength regime.

\subsection{Energy transport and phase structure}
The local energy transport is described by the time-averaged Poynting vector
\begin{equation}
\langle \bm{S}(\bm{r})\rangle=\frac{1}{2}\Re\left\{\bm{E}(\bm{r})\times \bm{H}^*(\bm{r})\right\}.
\label{eq:poynting}
\end{equation}
In the two-dimensional setting, $\langle \bm{S}\rangle$ is visualized as a vector field in the $x$--$z$ plane. We interpret the topology of this field in terms of streamlines, stagnation points, and vortical circulation. We also analyze phase maps $\phi(\bm{r})=\arg(E_\alpha)$ and identify phase singularities as points where a complex field component vanishes and the phase winds by $\pm 2\pi$ around a small loop \cite{NyeBerry}.

\section{Enhanced transmission through a subwavelength slit (TM)}
\subsection{Energy-flow reorganization across resonance}
\begin{figure}[ht]
\centering
\includegraphics[width=\linewidth]{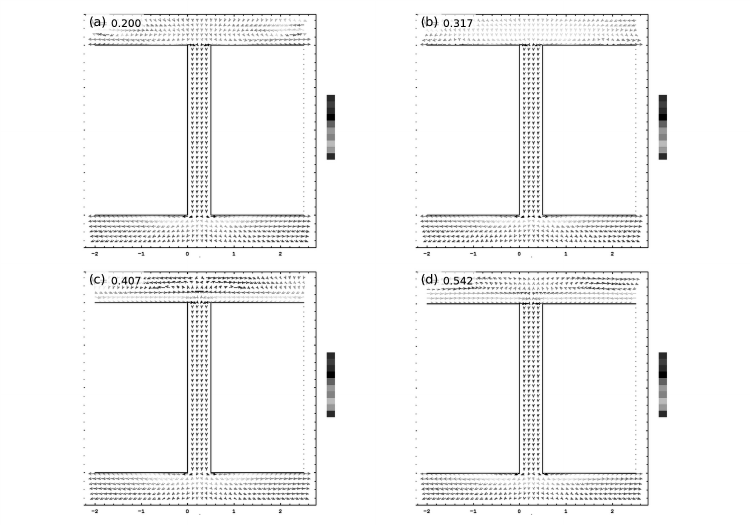}
\caption{Near-field energy flow through a subwavelength slit under TM illumination. Streamlines represent the time-averaged Poynting vector $\langle \bm{S}\rangle$ across a wavelength scan. The resonant regime exhibits pronounced energy funneling, localized backflow, and nontrivial streamline organization.}
\label{fig:slit_poynting}
\end{figure}

Figure~\ref{fig:slit_poynting} shows the evolution of the energy flow near a subwavelength slit under TM illumination as the wavelength is scanned through resonance. Away from resonance, streamlines remain comparatively smooth. Near resonance, energy is efficiently funneled into the slit and the flow becomes strongly structured, with localized recirculation and backflow regions near the aperture mouths. These features are the near-field signature of resonant excitation and interference of the dominant guided mode, consistent with the standard modal (Fabry--P\'erot) interpretation of enhanced transmission \cite{Takakura,BravoAbadPRB2004,LalannePRL}.

\subsection{Phase structure and singular features}
\begin{figure}[ht]
\centering
\includegraphics[width=\linewidth]{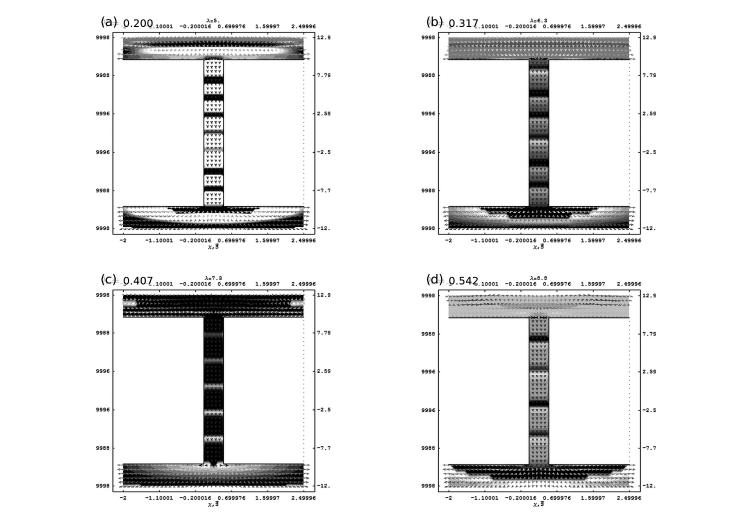}
\caption{Correlation between energy-flow topology and phase structure near a subwavelength slit. The resonant regime exhibits strong phase gradients and singular features associated with the excited modal field, which organize $\langle \bm{S}\rangle$ into vortical and saddle-type patterns.}
\label{fig:slit_poyphase}
\end{figure}

To connect the energy-flow structure with the complex near field, Fig.~\ref{fig:slit_poyphase} correlates Poynting streamlines with phase-related information. The resonant regime is accompanied by steep phase gradients and phase singularities (wavefront dislocations) \cite{NyeBerry,DennisProgOpt}. These singular features organize the local flow into vortical and saddle-type structures, providing an intuitive bridge between modal interference and energy-transport topology.

\subsection{Field localization and polarization selectivity}
\begin{figure}[ht]
\centering
\includegraphics[width=\linewidth]{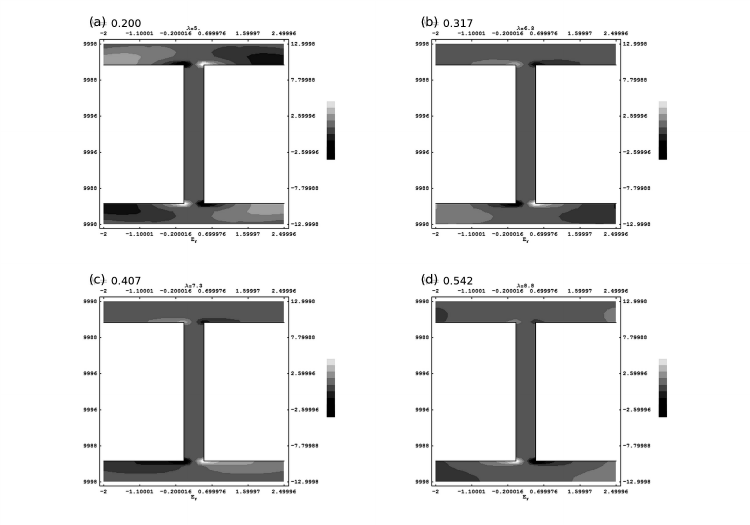}
\caption{Evolution of the transverse electric-field component $E_y$ under TM illumination. Resonant transmission is accompanied by strong localization near the slit edges, providing a direct near-field mechanism for polarization selectivity within the modal picture.}
\label{fig:slit_ey}
\end{figure}

Figure~\ref{fig:slit_ey} shows the evolution of the transverse electric-field component $E_y$ under TM illumination. At resonance, $E_y$ localizes strongly near the slit edges, consistent with excitation of the dominant TM mode and with charge accumulation at conducting boundaries. Below cutoff, TE modes are evanescent and do not sustain comparable buildup, explaining polarization selectivity.

\section{Enhanced transmission through a subwavelength channel}
We extend the analysis from a single slit to a finite-length subwavelength channel. Within modal theory, the channel supports TE and TM guided modes whose propagation constants depend on both width and length \cite{CollinBook,GarciaVidalRMP}. As in the slit geometry, the deep-subwavelength regime is dominated by the fundamental TM mode, while TE transmission is limited by cutoff and evanescence.

\subsection{Energy-flow evolution across resonance (TM)}
\begin{figure}[ht]
\centering
\includegraphics[width=\linewidth]{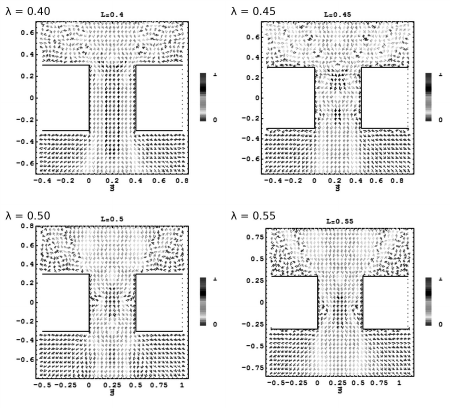}
\caption{Near-field energy flow in a subwavelength channel under TM illumination. The resonant regime again exhibits strong funneling and localized backflow/recirculation near the channel entrances, consistent with resonant excitation of the dominant TM guided mode.}
\label{fig:chan_poynting}
\end{figure}

Figure~\ref{fig:chan_poynting} shows the energy-flow field for a wavelength scan across the channel resonance. The key qualitative features observed for the slit persist: energy funneling into the channel, localized recirculation near the entrances, and backflow in the near field. The similarity between slit and channel demonstrates that the energy-flow topology is a robust consequence of modal excitation and interference rather than a geometry-specific artifact.

\subsection{Field evolution in the channel}
\begin{figure}[ht]
\centering
\includegraphics[width=\linewidth]{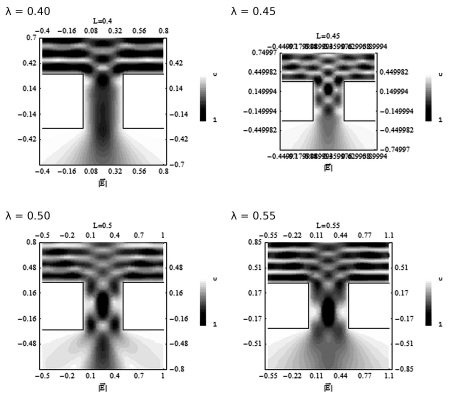}
\caption{Representative channel field evolution across the wavelength scan. Enhanced localization at resonance accompanies the energy-flow reorganization and is consistent with modal buildup inside the channel.}
\label{fig:chan_efield}
\end{figure}

Figure~\ref{fig:chan_efield} provides a complementary view of the channel field evolution across the same scan. Enhanced localization at resonance accompanies the energy-flow reorganization and is consistent with Fabry--P\'erot--like modal buildup inside the channel \cite{Takakura,GarciaVidalRMP}.

\subsection{TE polarization (contrast case)}
For TE polarization, the modal structure differs due to cutoff. Below cutoff, TE modes are evanescent and do not sustain strong resonant buildup; consequently, both transmission and near-field energy-flow structuring remain weak. TE results are therefore used as a contrast case and can be included as supplementary material.

\section{Discussion}
The resonant wavelengths and near-field distributions obtained here are consistent with standard modal and Fabry--P\'erot descriptions of subwavelength apertures \cite{Takakura,PortoPRL1999,GarciaVidalRMP}. The present results add a near-field energy-transport perspective: at resonance, the Poynting-vector field undergoes a qualitative reconfiguration that can be described in terms of streamline topology (vortices, saddle points, separatrices, and localized backflow). This interpretation naturally connects with the concept of optical currents and local energy transport discussed by Berry \cite{BerryEnergyFlow2009}.

Importantly, these signatures are observed in both slits and channels and do not rely on surface plasmon polaritons. The energy-flow topology therefore provides a unifying near-field language for enhanced transmission in perfect conductors and non-plasmonic regimes, complementing standard modal interpretations and offering a direct visualization of how energy is transported into and through deeply subwavelength apertures.

\section{Conclusion}
Using a TE/TM modal formulation, we analyzed enhanced transmission through subwavelength slits and channels via near-field energy transport. Resonant excitation of the dominant TM mode is accompanied by a pronounced reorganization of the time-averaged Poynting vector, including vortical and saddle-type structures and localized backflow. These features correlate with phase singularities in the near field and provide a unified, non-plasmonic interpretation of enhanced transmission in subwavelength apertures.

\section*{Funding}
(Insert funding information here.)

\section*{Acknowledgments}
(Insert acknowledgments here.)

\section*{Disclosures}
The authors declare no conflicts of interest.

\bibliographystyle{unsrtnat}
\bibliography{references}

\end{document}